\documentclass[journal]{IEEEtran}

\ifCLASSINFOpdf
\else
\usepackage[dvips]{graphicx}
\fi

\usepackage{cite}
\usepackage{amsmath}
\usepackage{amssymb} 
\usepackage{graphicx}

\begin{document}
\title{Block-level Double JPEG Compression Detection for Image Forgery Localization}

\author{Vinay Verma, Deepak Singh, and Nitin Khanna
\thanks{This material is based upon work partially supported by a grant from the Department of Science and Technology (DST), India (Award Number ECR/2015/000583)}
\thanks{Affiliation: Electrical Engineering at Indian Institute of Technology Gandhinagar, Palaj, Gandhinagar, Gujarat, India, 382355 (e-mail:\{vinay.verma, nitin.khanna\}@iitgn.ac.in, deepak.singh@msc2016.iitgn.ac.in)}
}
\maketitle

\begin{abstract}
Forged images have a ubiquitous presence in today's world due to ease of availability of image manipulation tools. 
In this letter, we propose a deep learning-based novel approach which utilizes the inherent relationship between DCT coefficient histograms and corresponding quantization step sizes to distinguish between original and forged regions in a JPEG image, based on the detection of single and double compressed blocks, without fully decompressing the JPEG image. 
We consider a diverse set of 1,120 quantization matrices collected in a recent study as compared to standard 100 quantization matrices for training, testing, and creating realistic forgeries. 
In particular, we carefully design the input to DenseNet with a specific combination of quantization step sizes and the respective histograms for a JPEG block. 
Using this input to learn the compression artifacts produces state-of-the-art results for the detection of single and double compressed blocks of sizes $256 \times 256$ and gives better results for smaller blocks of sizes $128 \times 128$ and $64 \times 64$.
Consequently, improved forgery localization performances are obtained on realistic forged images. Also, in the case of test blocks compressed with completely different quantization matrices as compared to matrices used in training, the proposed method outperforms the current state-of-the-art.  
\end{abstract}


\IEEEpeerreviewmaketitle
\section{Introduction}
\label{sec:djpeg_intro}
\IEEEPARstart{D}{igital} images are widely available due to maturing imaging technology and social media is filled with a large number of tampered digital images due to the easy availability of a wide range of image processing tools that do not require expertise to manipulate digital images. 
These manipulations done to the images are hard to detect with the naked eye.
Once tampered, these digital images can easily be shared to spread misinformation, such as causing severe damages to someone's reputation. 
An overview of various tools for authentication of digital images can be explored in~\cite{bib:farid2009_spm,bib:piva2013overview,bib:stamm2013overview}. 
Detection of the compression history of JPEG images is of great importance since most of the digital cameras and image processing tools use JPEG format to encode digital images and image compression history may indicate potential manipulations in a given JPEG image. 
A recent user study~\cite{bib:Park_2018_ECCV} collected 127,874 digital images for two years using a general forensic website, and found that 77.95\% images were in JPEG format, and out of this, 41.77\% were compressed using non-standard quantization matrices (Q-matrices). 
So, the development of a method to detect manipulations (if any) in a JPEG image, taking into account standard and non-standard Q-matrices, is essential.
In this letter, we address the problem of detecting single and double compressed JPEG blocks/patches (parts of an image aligned with $8\times8$ grid), which consequently helps us to locate the forged regions in a given JPEG image. 

There is a rich literature on methods related to detection of double compression in JPEG images using blocking artifacts in pixel domain and various handcrafted features of DCT coefficients in transform domain~\cite{bib:lukas03,bib:popescu04,bib:fu07,bib:luo2007icassp, bib:ye2007,bib:li08,bib:li2009passive, bib:lin2009fast,bib:farid2009exposing,bib:chen2011tifs, bib:bianchi2011improved, bib:amerini14,bib:wang2014,bib:taimori17,bib:pasquini14,bib:milani14}.
In terms of deep learning approaches~\cite{bib:wang2016double,bib:barni2017aligned,bib:verma2018dct,bib:li2017multi,bib:li2019doublejpeg}, Wang \textit{et al.}~\cite{bib:wang2016double} was first to use a one dimensional CNN with input as the histogram (range [-5,5]) of the DCT coefficients of first 9 frequencies, resulting in better performance than handcrafted features. 
Later, Barni \textit{et al.}~\cite{bib:barni2017aligned} used CNN with improved performances.
Classifying images among the uncompressed, single, and double compressed using CNN is addressed in~\cite{bib:amerini2017localization,bib:zeng2019detection}.
One drawback of all these systems, as mentioned earlier, is to consider only 100 standard Q-matrices, while there is a possibility of a much broader set of Q-matrices, as different camera manufacturers or algorithms are free to use their own Q-matrices. 
To address this issue, authors in~\cite{bib:Park_2018_ECCV} collected 1,170 unique Q-matrices (including the 100 standard ones), classified single and double compressed blocks using the histogram of de-quantized DCT coefficients as input to a CNN.  
Authors in~\cite{bib:Park_2018_ECCV} experimentally observed that appending the reshaped Q-matrix (quantization step sizes for all the 64 frequencies together), at the last three fully connected layers of their CNN improves the performance.

One step forward, we argue and propose a theoretically justified way of using the quantization step sizes (q-factors) with the corresponding histograms for each of the 64 frequencies individually, which is experimentally backed-up with improved performances.
For the histogram formation, we use quantized DCT coefficients directly extracted from JPEG bit-stream, unlike the method in~\cite{bib:Park_2018_ECCV}, which uses the de-quantized DCT coefficients (JPEG image is first decompressed and DCT is calculated to get the de-quantized DCT coefficients).
The process of decompression in this above procedure involves rounding and truncation operation, which can be avoided by directly utilizing the quantized DCT coefficients.
To summarize, following are the major contributions of this letter: 
\begin{enumerate}
	\item a new deep learning-based framework for detection of single and double compressed blocks, compressed with more diverse unique Q-matrices as compared to standard Q-matrices used in most of the methods except~\cite{bib:Park_2018_ECCV}. 
	\item a well-designed input, which is a specific combination of the histograms and corresponding q-factors, that is passed to DenseNet to learn the compression artifacts. 
	\item significant performance improvement for smaller size blocks, and in the real-life image forgery detection and localization capability.
	\item better generalization capability for the test blocks compressed with completely different Q-matrices.
\end{enumerate}

\section{Proposed System}
\label{sec:djpeg_prop_sys}
\subsection{Background and Terminologies}
\label{subsec:back_term}
JPEG compression of an RGB image involves first conversion to YCbCr space.
Each channel is divided into independent non-overlapping $8\times8$ blocks, shifted to signed integer range ([-128, 127]), and 2D Discrete Cosine Transform (DCT) coefficients are calculated. 
A quantization matrix (Q-matrix) of size $8\times8$ is used to quantize the corresponding DCT coefficients.
In general, two different Q-metrics, one for the Y channel and another one for the Cb and Cr channels are used. 
These quantized DCT coefficients are entropy encoded to get the JPEG bit-stream.
JPEG decompression, the process of getting the image back in the pixel domain, involves the reverse process, namely entropy decoding, dequantization, inverse DCT (IDCT), rounding, and truncation.

A JPEG image directly coming from a digital camera is single compressed (compression within camera module). 
Image manipulation involves first decompression, modification of some region of the image, and further re-saving in the JPEG format.
During the final compression, as detailed in~\cite{bib:lin2009fast}, the tampered $8 \times 8$ blocks become single compressed, while the untampered blocks become double compressed.
Thus, the detection of single and double compressed blocks (patches) in a given JPEG image helps to localize the tampered regions. 

\subsection{Design of Optimal Input}
\label{subsec:prop_approach}

In a JPEG patch, consider the collection of DCT coefficients at a frequency location $(u,v)$ ($u,v\in \left\{1,2,\ldots8\right\}$) to be $\{F(u,v)\}$.
For a single compressed patch, the collection of quantized DCT coefficients is $\left\{\left[\dfrac{F(u,v)}{Q_1(u,v)}\right]\right\}$, where $Q_1$ is the Q-matrix used for the first compression, $Q_1(u,v)$ is the quantization step size (q-factor) at $(u,v)$, and $[.]$ denotes rounding operation. 
Further, during decompression, de-quantization with $Q_1(u,v)$ and during the second compression, quantization with q-factor $Q_2(u,v)$ happens. 
So, for double compressed patch, neglecting rounding and truncation after IDCT operation, the collection of quantized DCT coefficients is $\left\{\left[\left[\dfrac{F(u,v)}{Q_1(u,v)}\right]\dfrac{Q_1(u,v)}{Q_2(u,v)}\right]\right\}$, exhibits double quantization artifacts with periodic peaks and valleys in the histograms for all $u,v\in \left\{1,2,\ldots8\right\}$.
Details of these periodic artifacts have been previously described in the literature~\cite{bib:lukas03,bib:popescu04,bib:lin2009fast}.

Quantized DCT coefficients for a JPEG patch can be directly obtained from the bit-stream of the JPEG file without fully decompressing using pysteg~\cite{bib:pysteg}. 
Irrespective of patch being single or double compressed, let $\left\{F_{q}(u,v)\right\}$ denote its collection of quantized DCT coefficients at a frequency location $(u,v)$.
Histogram of quantized DCT coefficients at $(u,v)$ with integer bins in the range $[-b , b]$, is defined as: $h(i) = |\{F_q(u,v) | F_q(u,v) = i\}|$, $i = [-b,b]$ ($|.|$ is cardinality of a set). 
Hence, histogram for a particular frequency $\mathbf{h}^{(u,v)}\in \mathbb{Z}^{(2b+1)}$. For all the 64 frequencies, the histogram is denoted as $\mathbf{H} = [\mathbf{h}^{(1,1)};\; \mathbf{h}^{(1,2)};\; \mathbf{h}^{(1,3)};\; \ldots\; \mathbf{h}^{(8,8)}]\in \mathbb{Z}^{64\times (2b+1)}$.

For a JPEG patch, it's header always contains the information of Q-matrix used during the final compression. 
We utilize the available q-factor for each of the frequency $(u,v)$ with their respective histograms in a specific way and use a CNN to learn to distinguish between single and double compressed JPEG patches.
For each frequency $(u,v)$, corresponding q-factor $Q(u,v)$ are repeated $(2b+1)$ times and channel-wise concatenated with the histogram $\mathbf{h}^{(u,v)}\in \mathbb{Z}^{2b+1}$ to get the feature representation of dimension $(2b+1) \times 2$.
For all frequencies, final feature representation corresponds to $\mathbf{X}$ having dimensions of $64\times (2b+1) \times 2$.
To summarize, for a JPEG patch, define a matrix of q-factors, $\mathbf{Q^{'}}$ as in Equation~\ref{eq:Q_dash}, and channel-wise concatenate the $\mathbf{H}$ with $\mathbf{Q^{'}}$ to get the final feature matrix $ \mathbf{X}\in \mathbb{Z}^{64\times (2b+1)\times 2}$. 
The rationale behind using the q-factor in this way is explained below.
In the case of single compression, the collection of quantized DCT coefficients \{$F_q(u,v)$\} for a frequency $(u,v)$, are dependent on the $Q_1(u,v)$ (q-factor at $(u,v)$) and is independent of other 63 q-factors of other frequencies. 
While for double compression \{$F_q(u,v)$\} is dependent on $Q_1(u,v)$ and $Q_2(u,v)$. 
Although for a double compressed patch, we do not have the access to first q-factor $Q_1(u,v)$, the relationship of q-factors with the corresponding quantized DCT coefficients is captured in the corresponding histogram.
Theoretical analysis of the relationship between the DCT coefficient's histogram with the q-factor can be found in~\cite{bib:lin2009fast}. 
The final feature representation $\mathbf{X}$ is fed as input to DenseNet~\cite{bib:huang2017densely} to learn to distinguish between single and double compressed JPEG patches.

\begin{align}
\label{eq:Q_dash}
\mathbf{Q^{'}} &= \begin{bmatrix}
Q{(1,1)} & Q{(1,1)}&\ldots& Q{(1,1)} \\
Q{(1,2)} & Q{(1,2)}&\ldots& Q{(1,2)} \\
\vdots & \vdots&\ddots& \vdots \\ 
Q{(8,8)}  &Q{(8,8)}&\ldots& Q{(8,8)}
\end{bmatrix}\in \mathbb{Z}^{64\times (2b+1)}
\end{align}

\subsection{Network Architecture}
\label{subsec:net_arch}

We utilized a variant DenseNet-121 architecture~\cite{bib:huang2017densely} for the two-class problem addressed here. 
Input to the network, $\mathbf{X}$, is passed through a convolutional layer (64 kernels, $7\times7$, stride = 2), followed by a max pooling layer ($3\times3$, stride = 2). 
Further, the network has four dense blocks with a growth rate of $k=32$ and the number of dense layers equal to 6, 12, 24, and 16, respectively.
Each dense layer consists of BN, ReLU, $1\times 1$ convolution ($4k$ kernels, stride = 1), BN, ReLU, $3\times 3$ convolution ($k$ kernels, stride = 1).
In a dense block, input to a dense layer is feature maps of all previous dense layers, and similarly, the output feature map of the current dense layer is used as input in the upcoming dense layers.
Each of the first three dense blocks are followed by three transition blocks consisting of BN, ReLU, $1\times 1$ convolution with compression factor 0.5, and an average pooling layer ($2\times 2$, stride = 2). 
Weights and biases of each convolutional layer are initialized with Xavier uniform initializer~\cite{bib:glorot2010understanding} and zero vector, respectively. 
The output of the last dense block is passed through a $7\times 7$ global average pooling layer. Finally, softmax is used to obtain the class probability scores. 
Categorical cross-entropy loss between the actual label and the predicted value is used as a loss function for optimization. 
Network is trained using Adam optimizer~\cite{bib:kingma2014adam} with the value of $\beta_1$ and $\beta_2$ set at their default values 0.9 and 0.999, respectively. 
Batch size is chosen to be 64 for 40 epochs. The initial learning rate for the first 30 epochs is set to 0.001 and then reduced to 0.0005 for the last 10 epochs.

\section{Experiments and Discussion}
\label{sec:djpeg_exp}

We have used the publicly available~\cite{bib:djpeg2018eccv} dataset by Park \textit{et al.}~\cite{bib:Park_2018_ECCV}. 
Dataset consists of nearly 1.14 million $256\times 256$ single and double compressed JPEG patches.
This is the only dataset that considers 1120 unique Q-matrices (including standard Q-matrices for quality factors 51 - 100) for creating single and double compressed patches.
These patches ($256\times 256$ ) extracted from 18,946 uncompressed images, captured from 15 different camera models were taken from RAISE~\cite{bib:dang2015raise}, Dresden~\cite{bib:gloe2010dresden}, and BOSS~\cite{bib:bas2011break}.
Single compressed blocks were created by compressing an uncompressed block with randomly chosen Q-matrix, while double compressed blocks were created by re-compression with another randomly chosen Q-matrix different from the first one. 
All the patches in this dataset~\cite{bib:Park_2018_ECCV}, and other images used in this letter are compressed by quantization of only Y channel.

Training parameters consist of histogram bin range controlled by $b$, train percentage, and patch-size, with their default values set to $b=100$, $30$\% , and $256 \times 256$, respectively.
Due to the time-complexity in training with $1.4$ million patches, we opted for two different train percentages, one with smaller number of training patches, where train, validation, and test set are 30\% (171,065 patches per class), 10\% (57,022 patches per class), and 10\% respectively. 
Using this split, model hyper-parameters are optimized on the validation set of 10\%. 
Another split uses a larger training set, 90/10 train/test split, with train and test sets as 90\% (513,194 patches per class) and 10\%, respectively. 
Test set of 10\% is kept fixed across all the experiments, and methods. 
During training and testing, single and double compressed classes have patches with the same image content but compressed different number of times.

To evaluate the performance of the systems, we use the metrics accuracy, True Positive Rate
(TPR), and True Negative Rate (TNR) defined as ${(tp+tn)}/{(tp+tn+fp+fn)}$, ${tp}/{(tp+fn)}$, and ${tn}/{(tn+fp)}$, respectively.
Where $tp$ and $tn$ denote the number of patches correctly predicted as double compressed and single compressed respectively. 
While $fp$ and $fn$ denotes the number of patches incorrectly predicted as double compressed and single compressed respectively.
Hence, TPR denotes the number of patches correctly detected as double compressed out of the total double compressed patches. 
Similarly, TNR denotes the number of patches correctly detected as single compressed out of the total single compressed.

\subsection{Histogram's Bin Variation}
\label{subsec:hist_bin_var}
The range of quantized DCT coefficients is image and frequency location dependent. 
Based on the initial experiments, we varied the range of histogram parameter $b\in\{60, 80,100,120\}$.
Results in Table~\ref{tab:hist_range} demonstrate a slight increase in the performance with the increased bin-range. 
Considering the trade-off among the accuracy, TPR, TNR, and the training time, the value of $b=100$ is fixed.
\begin{table}[!htb]
	\centering
	\caption{Performance comparison with different histogram bin ranges}
	\label{tab:hist_range}
	\begin{tabular}{|l|c|c|c|}
		\hline
		Bin Parameter & Test Accuracy (\%) & TPR (\%) & TNR (\%)\\ \hline
		b = 60          &  93.51       &  90.49   &   96.52  \\ \hline
		b = 80         &   93.59      &  89.69   &  97.48   \\ \hline
		b = 100        &    93.73     &   90.48  &  96.99   \\ \hline
		b = 120       &     93.74    &   90.20  &   97.28  \\ \hline
	\end{tabular}
\end{table}

\subsection{Comparison with State-of-the-Art Methods}
\label{subsec:res_analysis}
The performance of the proposed approach is compared with three methods~\cite{bib:Park_2018_ECCV}, Wang \textit{et al.}~\cite{bib:wang2016double} and Barni \textit{et al.}~\cite{bib:barni2017aligned}. 
As the latest paper~\cite{bib:Park_2018_ECCV} reported the accuracies for a 90/10 split, we have also used 90/10 split on the same dataset for a fair comparison. 
Table~\ref{tab:90_perf} shows the comparative performance of these methods. 
Due to the unavailability of exact 90/10 split used in~\cite{bib:Park_2018_ECCV}, for the proposed method, we used 5 different 90/10 train test random splits, and the mean test accuracy, TPR, and TNR of these 5 iterations with a standard deviation of 0.06\%, 0.20\%, 0.24\%, respectively, are reported. 
The (min, max) values of accuracies, TPR, and TNR for the proposed method are (94.40\%, 94.56\%), (91.52\%, 92.05\%), and (96.94\%, 97.48\%), respectively.
Results in Table~\ref{tab:90_perf} demonstrate the improvement in overall accuracy, ability to correctly classify single and double compressed patches as compared to other approaches.
Using RGB blocks directly as input to the DenseNet without the appropriate design of input, resulted in random 50\% accuracy, similar to~\cite{bib:Park_2018_ECCV}, with VGG-16~\cite{bib:simonyan2014very} network.

\begin{table}[!htb]
	\centering
	\caption{Performance of various approaches using 90\% patches for training}
	\label{tab:90_perf}
	\begin{tabular}{|c|c|c|c|}
		\hline
		Methods & Test Accuracy (\%) & TPR (\%) & TNR (\%)\\ \hline
		Wang \textit{et al.}~\cite{bib:wang2016double} & 73.05 & 67.74 & 78.37\\ \hline
		Barni \textit{et al.}~\cite{bib:barni2017aligned} & 83.47 & 77.47 & 89.43 \\ \hline
		Park \textit{et al.}~\cite{bib:Park_2018_ECCV} & 92.76 &90.90&94.59 \\ \hline
		Our & \textbf{94.49} &\textbf{91.74}&\textbf{97.25}\\ \hline
	\end{tabular}
	
\end{table}

\subsection{Patch-Size Variation and Effect of Using q-factors in the Proposed Approach} Performance of the proposed approach and approach in~\cite{bib:Park_2018_ECCV} with three different patch-sizes are shown in Table~\ref{tab:patch_perf}.
Our network with growth rate $k = 32$, has $\sim$6.9 million learn-able parameters as compared to the optimized CNN network provided by~\cite{bib:Park_2018_ECCV} which has $\sim$16.8 million learn-able parameters. The results of~\cite{bib:Park_2018_ECCV} are reproduced with 100 epochs, 64 batch-size, and learning $10^{-5}$.
Quantized DCT coefficients for the patch-sizes $64\times 64$ and $128 \times 128$ are extracted from the top left corner of the $256\times 256$ patches which also ensures the same number of training and testing patches across all the patch-sizes. 
Performance in Table~\ref{tab:patch_perf} emphasizes the advantages of the proposed way of combining q-factors with the respective histograms with overall improved performance as compared to the method in~\cite{bib:Park_2018_ECCV}.
The phrases ``without q-factors'' and ``with q-factors'' signify the input $\mathbf{X}$ with dimension ${64\times (2b+1)}$ and ${64\times (2b+1)} \times 2$, respectively. 
The performance gain as compared to the method in~\cite{bib:Park_2018_ECCV} is more significant with the smaller size patches, as the values of accuracy, TPR, and, TNR are improved significantly.
In `with q-factor' scenario as compared to `without q-factor', TPR, the number of correctly classified double compressed patches out of total double compressed patches, improves drastically, but TNR is marginally higher in `without q-factor' case.
\begin{table}[!htb]
	\centering
	\caption{Performance of proposed system with different patch-sizes $64\times 64$, $128\times 128$, and $256\times 256$ and effect of using q-factors in the proposed approach}
	\label{tab:patch_perf}
	\resizebox{0.45\textwidth}{!}{
		\begin{tabular}{|c|c|c|c|}
			\hline
			Methods   & Test Accuracy (\%) & TPR (\%) & TNR (\%)\\ \hline
			Park \textit{et al.}~\cite{bib:Park_2018_ECCV} ($64\times 64$)           & 83.94  & 76.64 & 91.24\\ \hline
			our ($64\times 64$) (without q-factors) & 85.19 & 76.26 & \textbf{94.12} \\ \hline
			our ($64\times 64$) (with q-factors) & \textbf{88.09} & \textbf{82.32} & 93.85 \\ \hline \hline
			Park \textit{et al.}~\cite{bib:Park_2018_ECCV} ($128\times 128$)            &  87.68& 82.78 & 92.58 \\ \hline
			our ($128\times 128$) (without q-factors)& 88.62 & 79.81 & \textbf{97.43} \\ \hline
			our ($128\times 128$) (with q-factors) & \textbf{91.26} & \textbf{86.84}& 95.68 \\ \hline \hline
			Park \textit{et al.}~\cite{bib:Park_2018_ECCV} ($256\times 256$)                &  90.30       & 84.89    & 95.70   \\ \hline
			our ($256\times 256$) (without q-factors)        &   91.23     &  85.13  &  \textbf{97.34}   \\ \hline
			our ($256\times 256$) (with q-factors) &    \textbf{93.73}     &   \textbf{90.48}  &  96.99  \\ \hline
		\end{tabular}
	}
\end{table}

\subsection{Forgery Localization: Quantitative and Visual Analysis}
\label{subsec:forg_local}

To quantify manipulation region detection, similar to~\cite{bib:Park_2018_ECCV}, we considered copy-move and blur manipulation type of forgeries. 
2100 TIFF images from the RAISE dataset~\cite{bib:dang2015raise} are randomly selected.
For each image, $1024 \times 1024 $ size smaller images are randomly cropped and compressed with a randomly chosen Q-matrix among the 1120 Q-matrices. 
Further, the same $1024 \times 1024 $ image is decompressed, and a random region of size $544 \times 544$ is chosen and pasted at another random location, and re-compressed with another randomly chosen Q-matrix which is different from the first Q-matrix.
A similar approach is used for creating blur manipulated images by applying a blur filter ($\sigma = 2$) to randomly chosen $544 \times 544$ patch. 
For ground-truth creation, two-step process is followed, 1) from any $8\times8$ block even if a single pixel is part of the tampered area, that block is labeled as tampered (single compressed), 2) then each $256\times256$ block with stride 32, is labeled as tampered (single compressed) if the majority of $8\times 8$ blocks inside it are tampered, otherwise untampered (double compressed). 
Table~\ref{tab:quant_measure}, shows the improved results with precision $P={tp}/{(tp+fp)}$, recall $R={tp}/{(tp+fn)}$, F-measure $F={2PR}/{(P+R)}$, for the detection of copy-move forgeries and blur manipulations.
\begin{table}[!htb]
	\centering
	\caption{Copy-move and blur manipulation detection}
	\label{tab:quant_measure}
	\begin{tabular}{|c|c|c|c|}
		\hline
		Methods    & Precision & Recall & F-measure \\ \hline
		Park \textit{et al.}~\cite{bib:Park_2018_ECCV} (copy-move)           &    0.7544       &  0.7871      &    0.7704       \\ \hline
		our  (copy-move) &      \textbf{0.7830}     &    \textbf{0.8161 }   &    \textbf{0.7992 }      \\ \hline
		Park \textit{et al.}~\cite{bib:Park_2018_ECCV} (Blurring)            &    0.6929      &    0.8005    &   0.7428        \\ \hline
		our (Blurring)  &   \textbf{0.7241}        &    \textbf{0.8322}    &     \textbf{0.7744 }     \\ \hline
	\end{tabular}
\end{table}

\begin{figure}[!htb]
	\centering
\includegraphics[width=0.46\textwidth]{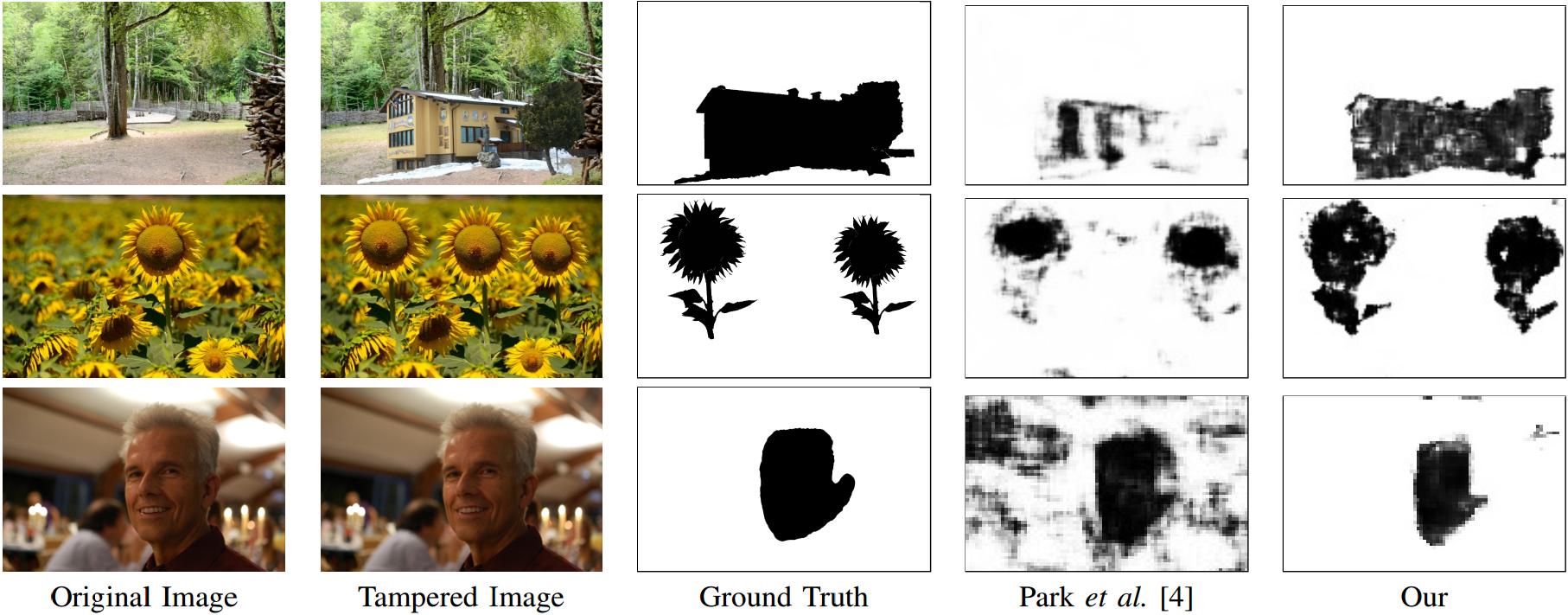}
	\caption{Examples of various types of forgeries and tampering localization. Top to bottom: splicing, copy-move, blurring}
	\label{fig:ex_tamp}
\end{figure}

For visualization, starting with images from RAISE~\cite{bib:dang2015raise}, six different types of manipulations~\cite{bib:Park_2018_ECCV} were performed in Adobe Photoshop, namely splicing, copy-move, color modification, content-aware object removal, blur manipulation, and object resizing, all using 1120 Q-matrices. 
Figure~\ref{fig:ex_tamp} shows the results of forgery localization for splicing, copy-move, and blur manipulation. 
For forgery localization in an image, the probability of blocks being double compressed $P(\hat{y}=1 \mid X)$ is shown with the stride set to 32 pixels.
Models used for quantitative and qualitative evaluations, were trained on the same 90/10 split (randomly chosen). 
The accuracy, TPR, and TNR for~\cite{bib:Park_2018_ECCV} was 92.26\%, 87.19\%, and 97.33\%, respectively, while for our method, accuracy, TPR, and TNR was 94.40\%, 91.87\%, and 96.94\%, respectively.

\begin{table}[!htb]
	\centering
	\caption{Performance for unseen Q-matrices}
	\label{tab:unseen_jpg}
	\begin{tabular}{|c|c|c|c|}
		\hline
		Methods     & Test Accuracy (\%) & TPR (\%) & TNR (\%)\\ \hline
		Park \textit{et al.}~\cite{bib:Park_2018_ECCV}          &   86.69    &  82.62  & 90.77    \\ \hline
		our         &    \textbf{92.83}     & \textbf{89.40}  &  \textbf{96.25}   \\ \hline
	\end{tabular}
\end{table}

\subsection{Analysis for Unseen Q-matrices}
\label{subsec:unseen_q}
In a real-life scenario, 1120 Q-matrices being more than standard Q-matrices, are still not a closed set, which implies need for a method that is able to handle unseen Q-matrices. 
In unseen Q-matrices cases, training and test patches have completely different Q-matrices.
We randomly picked 616 Q-matrices for training and the remaining 504 for testing, and the number of training and test patches was kept same as in the 30/10 split. The encouraging performances in Table~\ref{tab:unseen_jpg} demonstrate the applicability of the proposed approach as a universal single and double compressed block detector.

\section{Conclusions}
\label{sec:djpeg_conc}

In this letter, we have proposed a deep learning-based framework that combines the histogram of quantized DCT coefficients with the corresponding q-factors and utilizes DenseNet to extract the compression specific artifacts for classifying JPEG blocks/patches as single or double compressed. 
In a JPEG image, robust detection of single and double compressed blocks helps to localize the forged region (if any).
We showed significant improvement over the current state-of-the-art methods for the classification of JPEG blocks as single or double compressed, resulting in improved image forgery localization performances.
For smaller block sizes, the method performs significantly better and also generalizes for unseen compression scenarios. 
The limitation of the method, similar to existing ones, is the inability to classify patches compressed with the same Q-matrices in first and second compression.


\end{document}